\documentclass[12]{article}
\usepackage{graphics}

\begin{document}
\bibliographystyle{h-physrev}
\input{epsf}

\title{A Cohomological Interpretation of the Migdal-Makeenko Equations}
\author{A.Agarwal\thanks{abhishek@pas.rochester.edu} \\
\and
 S.G.Rajeev\thanks{rajeev@pas.rochester.edu} \\
University of Rochester. Dept of Physics and Astronomy. \\
Rochester. NY - 14627}
\maketitle

\begin{abstract}
The equations of motion of quantum  Yang - Mills theory (in the planar 
`large N' limit), when formulated 
in Loop-space are shown to have an anomalous term, which makes them 
analogous to the equations of motion of WZW models. The anomaly is 
the Jacobian of the  change of 
 variables from the usual ones i.e. the connection 
one form $A $, to the holonomy $U$. An infinite dimensional Lie algebra 
related to this change of variables (the Lie algebra of loop substitutions) 
is developed, and the anomaly  is interpreted as an element of the first 
cohomology of this Lie algebra. The Migdal-Makeenko equations are shown to 
be the condition for the invariance of the Yang-Mills generating 
functional $Z$ under the action of the generators of this Lie algebra. Connections of this formalism  to the 
collective field approach 
of Jevicki and Sakita are also discussed. 
\end{abstract}

\section{Introduction}
The purpose of the present paper is to investigate certain geometric and 
algebraic properties of the Migdal-Makeenko equations. What is of special 
interest is an anomaly contained in these equations,  
for which we provide a natural geometric interpretation.  To elaborate 
more on our motivation, we briefly review the loop-space formulation 
of 
QCD in the rest of this section.

 The natural gauge invariant variable for describing 
Yang-Mills theory, the Wilson loop ($W(\gamma )$), is defined as 
\begin{equation}
W(\gamma 
) 
= \left<tr U(\gamma )\right> = \left<tr P e^{\int_{\gamma }A} 
\right>\nonumber;
\end{equation} 
where, $\gamma $ is a loop in the base manifold of the principal 
bundle, and 
$U(\gamma )$  is the corresponding parallel transport operator. In 
geometric terms, changing variables from the connection one form $A$ to 
the parallel transport operator corresponding to loops $U(\gamma )$, 
amounts to talking in terms of differential forms on the loop space of a 
manifold instead of the connections on the manifold. Considerable 
literature has been devoted in physics\cite{Migdal1, Polyakov:glue, MME1, 
MME2, Loopspace, gambini1, gambini2, Yaffe1} towards understanding 
Yang-Mills 
using the language of loop spaces, and in mathematics\cite{chen1, chen2, 
chen3} towards constructing 
the theory of differential forms on loops spaces. The analogue of $A_\mu 
(x)$ in loop space is ${\cal{F}}\mu (\gamma (s))$, defined as;
\begin{equation}
{\cal{F }}_\mu (\gamma (s)) = \frac{\delta }{\delta \sigma _{\mu \nu 
}(\gamma (s))}U(\gamma )\dot{\gamma }_\nu (s) = U(\gamma _{0s})F_{\mu \nu 
}\dot{\gamma }_\nu (s)U(\gamma _{s0}).
\end{equation}
In the above equation $\frac{\delta }{\delta \sigma _{\mu \nu }(\gamma 
(s))}$ is the `area derivative' (in the sense of Migdal\cite{Migdal1, 
MME1, MME2}), $\gamma $ is a 
loop beginning and ending at $\gamma (0)$, and $U(\gamma _{0s}) (U(\gamma 
_{s0}))$ is the 
parallel transport operator associated with the path starting at $\gamma 
(0) (\gamma (s))$ and ending at $\gamma (s) (\gamma (0))$.\\
Classical Yang-Mills theory can now be recast in the language of ${\cal 
{F}}_\mu $, which is to be thought of as a one form in loop space. The 
classical equations of motion for Yang-Mills theory in this language are,
\begin{equation}
\frac{\delta }{\delta \gamma _\mu (s)}{\cal{F}}_\mu (\gamma (s)) = 0,
\end{equation} 
along with the identity,
\begin{equation}
\frac{\delta }{\delta \gamma _\nu (t)}{\cal{F}}_\mu (\gamma (s)) - 
\frac{\delta }{\delta \gamma _\mu (t)}{\cal{F}}_\nu (\gamma (s)) + 
\left[{\cal{F}}_\mu (\gamma (s)),{\cal{F}}_\nu (\gamma (t))\right] = 0.
\end{equation}
In the above equations, $\frac{\delta }{\delta \gamma _\mu (s)}$ is the 
usual variational derivative at $\gamma (s)$.\\
It was pointed out by Polyakov \cite{Polyakov:glue}, that these equations 
are 
very 
reminiscent of 
the 
classical equations of motion 
for chiral fields.\\

Interesting phenomena happen at the quantum level, where equation (3) no 
longer continues to be true as an equation for expectation values. The 
quantum analog of (3) (in the planar large N limit) is the Migdal-Makeenko 
equation (MME), which is,

\begin{equation}
\left< tr \frac{\delta }{\delta \gamma _\mu 
(s)}{\cal{F}}_\mu (\gamma(0))\right> = \frac{\partial 
^2 W(\gamma )}{\partial \gamma ^2_\mu (0)} =
-e^2\int\delta(\gamma (s) -\gamma (0))W(\gamma _{0s})W(\gamma
_{s0})\dot{\gamma _\mu}(0)\dot{\gamma _\mu }(s)ds
\end{equation}

In the equation above, the left hand side denotes the action of the 
loop-space Laplacian on the
Wilson loop. The Loop-space Laplacian \cite{Polyakov:glue, 
loopharmonic} is a second 
order differential
operator satisfying the Leibnitz rule, and is defined as follows.
\begin{equation}
\frac{\partial ^2}{\partial \gamma _\mu ^2 (s)} = \lim_{\epsilon \rightarrow 0}\int
_{-\epsilon }^{\epsilon }dt\frac{\delta ^2 }{\delta \gamma _\mu
(s + t/2)\delta \gamma _\mu (s - t/2)}
\end{equation}
The delta function on the right hand side is a
delta function on the underlying manifold, and it implies that the right
hand side of the equation is non-zero if the loop has a self intersection
at $\gamma (s) = \gamma (0)$. \\

Note: Since the Wilson loops themselves do not have any marked points, we 
can always identify the point at which the loop-Laplacian acts as the 
initial point of the loop. This convention leads to some notational 
simplifications, and hence we shall adhere to it in the rest of the 
paper.\\

This equation is usually derived by requiring the invariance of $W(\gamma 
)$ under a linear change of variables. In other words one considers the   
equation defining the Wilson loop, which is,
\begin{equation}
W(\gamma ) = \int D[A_\mu ] e^{\frac{1}{2e^2}\int tr F^2
dx}P\left(tre^{\int_{\gamma }A}\right),
\end{equation}
and  requires the invariance of the right hand side under $A_\mu
\rightarrow A_\mu + \delta A_\mu $. This change induces two effects; a
change in the action (proportional to $\nabla _\mu F_{\mu \nu })$, and a   
change in the holonomy associated with the loop $\gamma $, which involves 
the product of the holonomies associated with the closed loops $\gamma $  
splits into if it happens to have a self intersection. The cancellation of
these two changes generates the Migdal-Makeenko equation.\\

In the MME, the information about the Gauge theory is encoded in the
Loop-space Laplacian appearing in (4) because this  
operator
represents the change in the action under an infinitesimal variation of 
the
gauge field. The term that appears on the right hand side of the MME is a
sort of universal piece that is independent of the gauge theory that one  
might be studying. This universal term is a measure of the violation of 
the chiral equations of motion at the quantum level, its absence would 
lead us back to classical Yang-Mills theory. Hence it is  
necessary to understand the nature of this universal term, without which 
we can have no hope of understanding phenomena like confinement in 
quantum Yang-Mills theory.\\

In the present paper, we interpret this universal term as an anomaly 
\cite{Fujikawa} 
arising form the change in the Yang-Mills measure under a transformation 
that enables us to recast the theory directly in the language of objects 
defined on the loop space (e.g. $U(\gamma )$). Hence the structure of 
Yang-Mills theories in 
loop spaces is reminiscent of WZW models, with the right hand side of (5) 
representing the WZ term.\\

We accomplish the change of variables $A \rightarrow U(\gamma )$ through 
the action of some abstract Lie algebra generators $L(\gamma )$ on the 
connection one form; and we 
show that the MME is the condition for the invariance of the Yang-Mills 
generating functional $Z (= \int D[A\mu ] e ^{\frac{1}{2e^2} \int tr 
F^2})$ under the action of these generators, i.e. 
$L(\gamma ) Z = 0 \Rightarrow $ MME. We show further, that the expectation 
value of $L(\gamma )$ on the Yang-Mills measure produces the universal 
term present in the MME and we also observe that this anomalous term has a 
geometric meaning as an element of the first cohomology of this Lie 
algebra (the algebra of loop substitutions). We also prove that the effect 
of these $L(\gamma )$s on the gauge 
fixing and ghost part of the action is equivalent to a total BRS 
variation, which is an explanation for why no explicit gauge fixing is 
required in the MME.

\section{The Change of Variables, and the Lie Algebra of Loop 
Substitutions} 

The change in the Gauge field that we talked about in the preceding 
section, can be thought of as the action of  
 an abstract operator $L(\gamma )$(associated with the loop $\gamma $) 
on 
$A_\mu $, defined as,
\begin{equation}
L(\gamma )A_\mu(x) = -\epsilon\delta (x - \gamma (0))\dot{\gamma }_\mu 
(0)U(\gamma )
\end{equation}

The composition of two such generators is also  defined if the
two loops have a space-time point in common:

\begin{equation}
L(\gamma _2)L(\gamma _1) = -L(\gamma _{2,0t} \circ
\gamma_{1} \circ \gamma
_{2,t0})\dot{\gamma _2}(t)\dot{\gamma _1}(0)\delta(\gamma_2(t) -
\gamma_1(0))dt
\end{equation}

The composition above is well defined if $\gamma _2 (t) =  \gamma _1   
(0) $ for some value of t. When the condition is satisfied then the
product of the generators corresponds the generator associated with
the composed curve starting at $\gamma_2 (0)$,
going up to the point $\gamma _1(0)$ along $\gamma _2$, then coming back  
to $\gamma _1(0)$ by going around $\gamma _1$, and then back to $\gamma   
_2 (0)$ along $\gamma _2$.

So these abstract generators form an algebra. In-fact, they form a Lie 
Algebra (a generalization of the free Lie algebra \cite{Rajeevwilson}) 
with the following 
lie bracket:
\begin{eqnarray}
[L(\gamma_2),L(\gamma_1)] &=& \int _0 ^1dt[L(\gamma_{1,0t}\circ
\gamma_{2} \circ
\gamma_{1,t0})\dot{\gamma _1}(t)\dot{\gamma _2}(0)\delta (\gamma _1(t) -
\gamma _2(0)) - \nonumber\\
& &L(\gamma _{2,0t}
\circ \gamma _{1} \circ \gamma _{2,t0})\dot{\gamma
_2}(t)\dot{\gamma
_1}(0)\delta (\gamma _2(t) - \gamma _1(0))]
\end{eqnarray}

It is easy to check that given the above Lie bracket,the Jacobi identity 
is 
also satisfied by the 
generators.

The generators have a natural action on the Wilson Loops, which is given 
below.
\begin{equation}
L(\gamma _1)W(\gamma ) = \int _0 ^1 dt W(\gamma _{0t}\circ \gamma_1\circ 
\gamma_{t0})\delta (\gamma (t) - \gamma _1(0))\dot{\gamma }_\mu 
(t)\dot{\gamma }_{1\mu }(0)dt
\end{equation}

\section{Migdal-Makeenko Equation Revisited}
Given the Yang-Mills generating functional, we now show that, the 
Migdal-Makeenko equation is equivalent to the condition $L(\gamma )Z = 0$, 
and the universal term in the loop-space equations is the expectation value 
of the change in the measure of integration in $Z$ under the action of 
$L(\gamma )$.\\

The change of variables we are concerned with here is,
\begin{equation}
A^{ a}_{\mu b}(\gamma (0)) \rightarrow A^{\prime a}_{\mu b}(\gamma (0)) = 
A^a_{\mu b}(\gamma (0) - \epsilon \left(L(\gamma )A^a_{\mu 
b}(\gamma(0))\right) = 
A^a_{\mu b}(\gamma (0)) + \epsilon \dot{\gamma }_\mu (0)U(\gamma )^a_b
\end{equation} 
Or to express the same thing in words, only the gauge field at the 
space-time point corresponding to the initial point of the loop $\gamma $ 
undergoes a change proportional to the parallel transport operator 
associated with the loop $\gamma $. The gauge fields at all the other 
space-time points are left unchanged. Now; 
\begin{eqnarray} 
L(\gamma )Z & &= \int \left( L(\gamma )D[A\mu 
]\right)e^{\left(\frac{1}{2e^2}\int tr F^2 dx\right)} + \left< L(\gamma 
)\frac{1}{2e^2}\int tr F^2 dx \right> \nonumber \\
& &= \left<det J - 1\right> + \left<L(\gamma )\frac{1}{2e^2}\int tr 
F^2dx\right>,
\end{eqnarray}
Where $J$ is the Jacobian corresponding to the change of variables (11). 
The 
Jacobian can be written explicitly as follows,
\begin{eqnarray}
J^{a c \gamma (0) \nu }_{b d x \mu} & &= \frac{\delta A^{\prime a} _{\mu b 
}(\gamma (0))}{\delta A^{d}_{\nu c}(x)}  \\
& &= \delta (\gamma (0) - 
x)\delta 
^\nu _\mu \delta ^a _d \delta ^c _b + \epsilon \dot{\gamma }_\mu (0) \int 
_0 ^1 dt U(\gamma _{0t})^a _d \delta (\gamma (t) - x )U(\gamma _{t0})^c _b 
\dot{\gamma }_\nu (t)dt \nonumber
\end{eqnarray}

In deriving (14), we have used the relation,

\begin{equation}
\frac{\delta A^{a}_{\mu , b}(x)}{\delta A^d _{\nu , c}(y)} = \delta
(x - y)\delta ^{\nu }_{\mu }\left( \delta ^a_d \delta ^c_b +
\frac{1}{N}\delta ^a _b \delta ^d_c \right),
\end{equation}
and have dropped the $O(1/N)$ terms.\\

Suppressing the color, Lorentz and space-time indices, it is possible to 
think of $J$ as a tensor, i.e. $J = 1 + \epsilon R$, where $R$ corresponds 
to the tensor in (13) that involves the product of the two parallel 
transport operators.\\
Hence,
\begin{eqnarray}
\left< det J - 1\right> & &= \left< e^{trlog J} -1 \right> = -\epsilon 
\left< 
tr R\right> = -\epsilon \left< R^{a b \mu \gamma (0)}_{b a \mu \gamma 
(0)}\right> \\
& &= -\epsilon \int_0 ^1 W(\gamma _{0t})W(\gamma _{t0})\delta 
(\gamma (t) -\gamma (0))\dot{\gamma }_\mu (t) \dot{\gamma }_\mu (0)dt, 
\nonumber
\end{eqnarray}
which is nothing but the universal term in the MME. This calculation shows 
that it is indeed an anomaly (in the sense of Fujikawa \cite{Fujikawa}) 
because it arises 
from the change in the measure of integration under a non-linear change of 
variables.\\

The second term in (13) can be evaluated using the action of 
$L(\gamma )$ on  the gauge field, as given in (12). It follows easily 
that,
\begin{equation}
\left<L(\gamma )\frac{1}{2e^2}\int tr F^2 dx \right> = -\frac{\epsilon 
}{e^2}\left<tr(\nabla _\mu F_{\mu \nu }(\gamma (0))\dot{\gamma }_\nu 
(0)U(\gamma ))\right> = -\frac{\epsilon }{e^2} \frac{\partial ^2}{\partial 
\gamma ^2 _\mu (0)}W(\gamma )
\end{equation}

So substituting the results of equations (16) and (17) back in (13), we 
see that 
the Migdal Makeenko 
equation is the condition for the invariance of the Yang-Mills generating 
functional under the action of the generators of loop substitutions.

\section{Cohomology}

In this section we shall try to extract the geometric meaning of the 
anomaly in the Loop-space equations. The Migdal-Makeenko equations 
represent the invariance of the Yang-Mills generating functional under the 
action of the Lie algebra generators $L(\gamma )$. One may regard these 
generators as vector fields, and think of the anomalous term as resulting 
from the action of an abstract one form $\omega $ (the anomaly one form)  
 on these generators. Or 
more explicitly,
\begin{equation}
\omega (L(\gamma )) = \int _0 ^1dt W(\gamma _{0t})W(\gamma _{t0})\delta 
(\gamma (0) - \gamma (t))\dot{\gamma }_\mu (t) \dot{\gamma }_{\mu }(0). 
\end{equation}

We claim that $\gamma $ is an element of the first cohomology of the lie 
algebra described above. We first note that $\omega $ is closed, i.e. 
$d\omega = 0$. Or in other words;
\begin{equation}
L(\gamma _2 )\omega (L(\gamma _1)) - L(\gamma _1)\omega (L(\gamma _2)) = 
\omega \left([L(\gamma _2),L(\gamma _1)]\right)
\end{equation}
{\bf Proof:}\\
We start with the right hand side of (19).
\begin{eqnarray}
&&\omega \left([L(\gamma _2),L(\gamma_1 )]\right) = \nonumber \\
&&\hbox{ }\omega \left( \int_0 ^1 
ds L(\gamma_{1,0s}\circ\gamma _2 \circ \gamma _{1,s0})\delta (\gamma 
_2(0) - \gamma _1(s))\dot{\gamma }_{1\mu }(s)\dot{\gamma }_{2\mu 
}(0)ds\right) - \left(\gamma _1 \rightarrow \gamma _2 \right)\nonumber\\
&&\hbox{ }
\end{eqnarray}
Now looking at the first term on the right hand side of the equation above 
we notice that 
$\omega $ is being evaluated on the generator corresponding to $\gamma 
_{1,0s}\circ \gamma _2 \circ \gamma _{1,0s}$; assuming that $\gamma _1(s) 
= 
\gamma _2(0)$, for some $s$, (this has to be true for the equation to be 
non-trivial).  So by the definition of $\omega $, this will be non-zero if 
the composed curve self intersects itself at $\gamma _1(0)$. Hence we 
require that either $\gamma _1(t) = \gamma _1(0)$, or that 
$\gamma _2(t) = \gamma _1(0)$ for some value of the parameter $t$. It is 
easy to see that the product of Wilson loops obtained when the second 
condition holds good is symmetric under $\gamma _1 \rightarrow \gamma 
_2$and hence does not contribute to the commutator. Hence we shall focus 
only on the first case, i.e. $\gamma _1(t) = \gamma _1(0)$. Now depending 
on whether $t<s $ or $t>s$, we have;
\begin{eqnarray}
&&\omega \left([L(\gamma _2),L(\gamma _1)]\right) = \nonumber\\
&&\mbox{ }\int_0^1\,dt\int_0^1\,ds \left(\delta(\gamma _(t) - \gamma _1(0))\delta 
(\gamma _1(s) - 
\gamma _2(0)) )\dot{\gamma }_{1\mu }(t)\dot{\gamma }_{1\mu }(0)
\dot{\gamma }_{1\nu }(s)\dot{\gamma }_{2\nu
}(0) \right)\nonumber\\
&&(W(\gamma _{1,0t})W(\gamma _{1,ts}\circ \gamma _2 \circ \gamma 
_{1,st})\theta (t<s) +\nonumber\\
& & W(\gamma _{1,0s}\circ \gamma _2 \circ \gamma _{1,s0})W(\gamma 
_{1,0t})\theta 
(t>s) ) - \nonumber \\
& &(\gamma _1 \rightarrow \gamma _2). 
\end{eqnarray}
Now we see that the left hand side of (19) is 
\begin{eqnarray}
&&L(\gamma _2)\omega (L(\gamma _1)) - (\gamma_1 \rightarrow \gamma _2 ) =\nonumber\\ 
&&\mbox{ }\int _0 ^1 dt (\delta (\gamma _1(t) - \gamma _1(0))\dot{\gamma }_{1\nu 
}(t)\dot{\gamma }_{1\nu }(0) \nonumber\\
& &(\left( L(\gamma _2)W (\gamma 
_{1,0t})\right)W(\gamma _{1,t0}) + W(\gamma 
_{1,0t})\left( L(\gamma _2)W(\gamma _{1,t0})\right)) - \nonumber\\
& &(\gamma _1 \rightarrow \gamma _2)
\end{eqnarray}
Now using the natural action of the generators on the Wilson loops (as 
given in (10)), we obtain,

\begin{eqnarray}
&&L(\gamma _2)\omega (L(\gamma _1)) - (\gamma_1 \rightarrow \gamma _2 ) =\nonumber\\
&&\mbox{ }\int _0^1dt\int
_0^1ds \left(\delta(\gamma _(t) - \gamma _1(0))\delta (\gamma _1(s) -
\gamma _2(0)) )\dot{\gamma }_{1\mu }(t)\dot{\gamma }_{1\mu }(0)
\dot{\gamma }_{1\nu }(s)\dot{\gamma }_{2\nu
}(0) \right) \nonumber\\
& &(W(\gamma _{1,0t})W(\gamma _{1,ts}\circ \gamma _2 \circ \gamma
_{1,st})\theta (t<s) +\nonumber\\
& & W(\gamma _{1,0s}\circ \gamma _2 \circ \gamma _{1,s0})W(\gamma
_{1,0t})\theta
(t>s) ) - \nonumber \\
& &(\gamma _1 \rightarrow \gamma _2).
\end{eqnarray}
Noting the equality of the right hand sides of (21) and (22), we conclude 
that $d\omega = 0$. \\ Even without constructing an explicit proof, it is easy to convince oneself that 
$\omega $ cannot be exact because of the presence of the logarithm in its 
definition; $\omega = <\exp (tr 
\log J) -1>$.\\Hence the one form $\omega $ related to the anomalous 
term in the Migdal-Makeenko equation is an element of the first cohomology 
of the Lie algebra of Loop substitutions. \\

{\bf Comments on the Connection to the  collective field formalism:}This approach towards understanding the 
loop 
equations is closely tied to previous advances made  by Jevicki and Sakita 
\cite{Collective1, Collective2, 
Collective3, Collective4} towards formulating Yang-Mills theory on loop 
spaces. The loop space theory in their approach was defined by the generating functional $(Z_{JS})$ defined 
by\cite{Collective4};
\begin{equation}
Z_{JS} = \int [DW{\gamma }] e^{-S + \log J_{JS}},
\end{equation}
Where $ J _{JS}$ is the Jacobian arising from transforming to loop space variables. This Jacobian is different 
from the ones considered in the present paper because the changes of variables considered in the collective 
field approach are not necessarily infinitesimal. This Jacobian was formally shown to satisfy the following 
equation.
\begin{equation}
\frac{\delta \log J_{JS}}{\delta W(\gamma )} = -\Sigma _{\gamma '}\Omega ^{-1}(\gamma , \gamma ')w(\gamma '),
\end{equation}
where,
\begin{equation}
\Omega (\gamma, \gamma ') = \int d^4x\frac{\delta W(\gamma )}{\delta A^a_\mu (x)}\frac{W(\gamma ')}{\delta 
A^a_\mu (x)},
\end{equation}
and,
\begin{equation}
w(\gamma ) = -\int d^4 x \frac{\delta ^2 W(\gamma )}{\delta A_\mu (x)^2}.
\end{equation}
The Migdal - Makeenko equations can now be thought of as the condition for the invariance of $Z_{JS}$ under 
the action of $L(\gamma )$. Or in more precise terms, the relation between the Jacobian appearing in the 
collective field formalism and the anomaly one form may be expressed formally as the following equation,
\begin{equation}
\omega (L(\gamma )) = -<L(\gamma )\log J _{JS}>.
\end{equation} 
Although the precise definition of this Jacobian in the continuum remains a reasonably open question, 
(though it has 
been understood in simpler lattice like theories \cite{Collective3, 
entropy, Yaffe1}), its deformation under the 
action of the algebra introduced 
in this paper is a much better behaved quantity (namely the anomaly one form ), for which we now have a 
cohomological interpretation.

\section{Loop substitution generator, and BRS invariance}
It is well known that the loop space equations do not require explicit 
gauge fixing. 
In this final section we realize this very desirable feature of the loop 
equations in the formalism set up in this paper by showing  that the 
effect of the generator 
of loop-substitution on the Gauge-fixing and Ghost part of the action is 
 a total BRS variation.\\

The gauge fixed form of the Yang-Mills action is,
\begin{equation}
S = \int \left[-\frac{1}{4}tr F^2 -\frac{1}{2\xi }tr(\partial ^{\mu }A_\mu 
)^2 -i tr(\partial ^\mu \bar{C})(\nabla _\mu C)\right].
\end{equation}
The action is invariant under the BRS transformations,

\begin{eqnarray}
& &\delta A = \lambda \nabla _\mu C ,\\
& & \delta C = \frac {i\lambda }{2}g[\bar{C},C]_+, \delta \bar{C} = 
-i\lambda \frac{1}{\xi }\partial ^\mu A_\mu,
\end{eqnarray}
where, $\lambda $ is the BRS parameter.\\

It is easy to see that,
\begin{equation}
\delta U(\gamma ) = ig\lambda [C(\gamma (0)),U(\gamma )].
\end{equation}
Now using the definition of the action of $L(\gamma )$, we see that,
\begin{eqnarray}
&&-i\lambda L(\gamma )(S_{Gauge-Fixing} + S_{Ghost}) = \nonumber\\
&&\hbox{ }i\lambda L(\gamma 
)\int 
\left[ \frac{1}{2\xi }tr(\partial ^{\mu 
}A_\mu
)^2 +i tr(\partial ^\mu \bar{C})(\nabla _\mu C)\right] =\nonumber \\
&&\hbox{ }-i\lambda tr 
\left[ \frac{1}{\xi }\partial ^\nu [\partial ^\mu A_\mu(\gamma 
(0))]U(\gamma ) + g[\partial _\nu \bar{C}(\gamma (0))][C(\gamma (0)), 
U(\gamma )]\right]\dot{\gamma }_\nu (0) =\nonumber \\
&&\hbox{ }\delta tr\left(\partial ^\nu \bar{C}(\gamma (0))U(\gamma 
)\right)\dot{\gamma }_\nu (0),
\end{eqnarray}
which is a total BRS variation.

{\bf Acknowledgment:}  We would like to thank G. Krishnaswami, 
L. Akant, T. Kohno and A. Constandache for useful discussions. This work 
was supported in 
part by the US 
Department of
Energy, Grant No. DE-FG02-91ER40685

\bibliography{abhishekbib}

\end{document}